Indexing Electron Backscatter Diffraction Patterns with a Refined Template Matching Approach

**A. Foden[1*], D.M. Collins[2,3], A. J. Wilkinson[2], T. B. Britton[1]**

1. Department of Materials, Imperial College London, SW7 2AZ

2. Department of Materials, University of Oxford, OX1 3PH

3. School of Metallurgy and Materials, University of Birmingham, B15 2TT

*email: a.foden16@imperial.ac.uk; t: @AlexFoden

Graphical Abstract

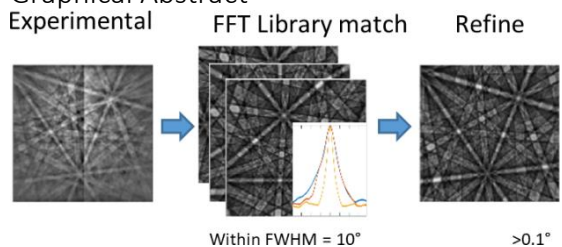

Abstract

Electron backscatter diffraction (EBSD) is a well-established method of characterisation for crystalline materials. Using this technique, we can rapidly acquire and index diffraction patterns to provide phase and orientation information about the crystals on the material surface. The conventional analysis method uses signal processing based on a Hough/Radon transform to index each diffraction pattern. This method is limited to the analysis of simple geometric features and ignores subtle characteristics of diffraction patterns, such as variations in relative band intensities. A second method, developed to address the shortcomings of the Hough/Radon transform, is based on template matching of a test experimental pattern with a large library of potential patterns. In the present work, the template matching approach has been refined with a new cross correlation function that allows for a smaller library and enables a dramatic speed up in pattern indexing. Refinement of the indexed orientation is performed with a follow-up step to allow for small alterations to the best match from the library search. The refined template matching approach is shown to be comparable in accuracy, precision and sensitivity to the Hough based method, even exceeding it in some cases, via the use of simulations and experimental data collected from a silicon single crystal and a deformed α-iron sample. The speed up and pattern refinement approaches should increase the widespread utility of pattern matching approaches.

Keywords
Electron Microscopy; Computer vision; Image processing

Introduction

Electron backscatter diffraction (EBSD) enables the rapid acquisition of the phase and orientation of the unit cells near the surface of a crystalline material [1]. It is used widely in the materials, mechanical engineering, and geoscience communities. In a scanning electron microscope, a high voltage (typically 5-30 keV) electron beam is scanned over a highly tilted sample and Kikuchi patterns, hereafter referred to as electron backscatter patterns (EBSPs), are generated. These patterns contain bands of raised intensity, called Kikuchi bands, generated due to diffraction from lattice planes in the crystal. Each pattern contains significant information concerning the lattice parameters (and deviations due to lattice strain), atomic distribution within the lattice, crystal phase, and the crystal orientation [2]. With the addition of a phosphor screen and CCD camera, the process of capturing EBSPs became automated [3–5], and the introduction of the Hough transform provided rapid processing of the patterns to determine crystal orientations [6,7]. This Hough transform (a subset of the Radon transform) converts the lines found in the EBSP to points within Hough space. Each point in Hough space represents the length tangential to each line from a set point (ρ) and the angle from the horizontal to that line (θ). Due to the invariance of interplanar angles with rotation, and that the EBSP is a direct projection of the crystal planes, a look up table of interplanar angles can be generated and used to index the bands found in Hough space (for more information, please see the recent release of an open source indexing approach [8]). The Hough transform is well suited to indexing EBSPs, however, it is limited in its application to the location of simple linear geometrical features and crystal structures. In practice, the accuracy and precision of the Hough transform method are based on how well the points can be identified in Hough space; typically measuring only the location of band centres and ignoring the relative intensity and width of the bands [9].

One of the strategies to improve the Hough based indexing method is template matching, e.g. the "Dictionary Indexing" approach [10,11]. In this method, reference patterns are taken from the 2D intensity distribution within EBSPs derived from high quality dynamical simulations. Patterns are calculated for select orientations to form a large library (or dictionary) of potential templates. In a captured map, the experimental patterns are sequentially template matched (i.e. compared) with the template library and the closest match is taken to be orientation of the unit cell. Matching in the Dictionary Indexing approach is performed via a pixel by pixel comparison, with the "normalised dot product" (NDP) and the orientation precision is ultimately limited to the sampling of the orientation space used to generate the library and therefore the library size. It follows that a very large library is required if a good orientation resolution is required. Further, the relatively narrow width of the local peak in the NDP as a function of misorientation from the true crystal orientation sets a minimum requirement on the sampling (such that a close enough template can be found) which itself leads to a substantial library size. This limits wide scale implementation due to the high computational costs associated with generating, storing, and matching with the library.

In this manuscript, we introduce a refined template matching approach. This utilises a Fourier domain based pattern matching method [12] and a fast refinement of the orientation, from which we demonstrate several advantages:

1. A wider matching peak than the NDP is attained. This enables comparison with a sparser template library (i.e. fewer patterns within the library).
2. The fast Fourier transform (FFT) based cross correlation function (XCF) enables whole pattern translations to be accounted for. These can accommodate large changes and/or uncertainties in the pattern centre (PC) as well as orientation changes.
3. A refinement of the orientation can be rapidly calculated to provide an orientation precision which is substantially better than the precision of library sampling.

To be more precise, in traditional template matching, we usually generate a library ($\boldsymbol{M}_i^k$) of $k$ potential templates and compare these with our experimental observation. The library is generated to populate orientation space with finely-spaced sampling. For the Dictionary Indexing method, each experimental pattern ($E_i$) is compared with each $k^{\text{th}}$ pattern from the library (in turn) via the average inner product, where each pattern is written as a linear vector of pixels:

$$C^k = \frac{1}{d}\sum_{i=1}^{d} \frac{\boldsymbol{M}_i^k \boldsymbol{E}_i}{|\boldsymbol{M}_i^k||\boldsymbol{E}_i|} \quad \text{Equation 1}$$

The objective of this comparison is to find subsequently the best match, i.e. highest value of $C^k$, where each experimental pattern and the respective library pattern are most similar.

The core idea of this paper is that the inner product is related to a special case (the point at the origin) of the cross correlation image, which contains information about the (translational) shift in each pixel ($m, n$) between two image arrays, such that:

$$C_{m,n}^k = \sum_{i,j} M_{i+m,j+n}^k E_{i,j}. \quad \text{Equation 2}$$

The information about these shifts can be extracted within the frequency domain via the FFT which is computationally inexpensive and enables frequency based pattern filtering.

$$C_{m,n}^k = \text{iFFT}(\text{FFT}M^k \odot \text{FFT}E) \quad \text{Equation 3}$$

We demonstrate that by including the shifts as a degree of freedom, we can broaden the sampling of orientation space and normalise the cross correlation more formally. The additional shift information can then be used in multiple ways to improve the matching procedure, details are given in later sections of the paper.

Method Overview

Figure 1 outlines the template matching methodology, which we introduce briefly here. Each step is described in detail subsequently.

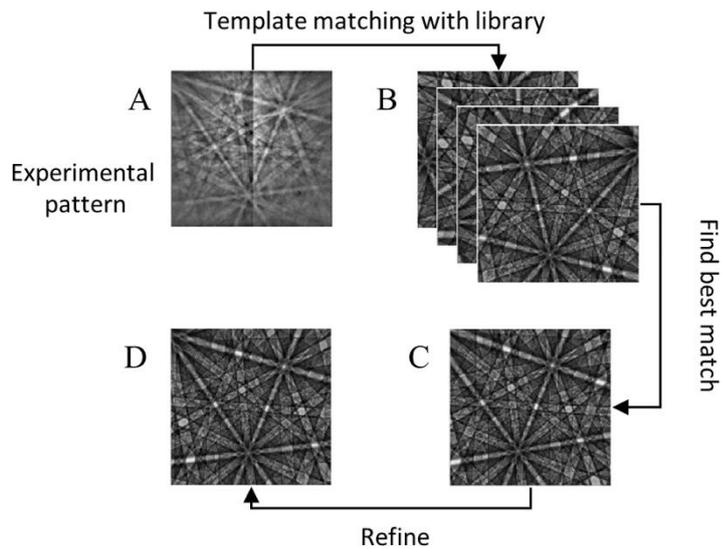

*Figure 1: The template matching method follows the following steps: A) An experimental pattern is imported for indexing. B) A library of templates is searched using an XCF to determine the orientation in SO(3) space (the 3D-rotation group) closest to the experimental pattern (C). D) The best orientation is refined to match the experimental orientation.*

Firstly, we introduce the method of templated matching, based upon the NDP (fixed templates) and 2D-FFT correlation (template translation allowed) approaches.

Next we introduce the generation of a library of EBSPs (templates), generated using a uniform sampling of the 3D-rotation group (hereon denoted as SO(3) space), with the MTEX 5.0.1 package for MATLAB [13]. The patterns in the library are compared to a pattern to be indexed using an image XCF. The crystal orientation of the test pattern is defined from the pattern with the closest match from the library. For the comparison, EBSPs are read in and background corrected, the patterns are then cropped to be square (optimal for a 2D FFT).

The discrete (and for the 2D-FFT case relatively sparse) nature of the library has prompted us to create a refinement step to interpolate from the best templated orientation to a more precise ultimate orientation. The refinement step approach also enables the pattern centre of the library and the experiment to be different and subsequently corrected (which is important for large area maps). The refinement is applied to the best match from the library search, comprising a translation in the plane of the detector (x and y translations, after correcting for pattern centre changes across the map) and a rotation about the normal to the detector (z - from a log-polar transform); this is iterated until a sufficiently good match is found and the final orientation is calculated.

Image Cross Correlations
Pattern matching can be performed using either the NDP [14–16], or alternatively the 2D-FFT [17–20].

Normalised Dot Product
The Dictionary Indexing method of pattern matching uses the inner product, which is also called the normalised dot product (NDP) [10]. This is calculated using Equation 1. Chen et al. [10] introduces the Dictionary Indexing method and comments that this index is used as a ranking to determine the best match from the dictionary of patterns. The inner product is calculated to have a value of 1, when the intensity distributions in both patterns are equal.

It is not clear what value this should be for two patterns with a similar EBSP, and therefore crystal, structure but different intensity values. Furthermore, Chen et al. [10] suggests only a SVD based background correction and they do not suggest significant image processing (e.g. contrast normalisation) which is typically performed for cross correlation and this may result in difficulties in evaluating the relative quality of the matching.

For the present work, we utilise contrast normalisation (as the intensity distribution within the experimental pattern is a near arbitrary unit) and utilise software based background correction for our experimental patterns. With the NDP, two identical images have $r_{fg} = 1$, and two completely dissimilar images have $r_{fg} = 0$. Unfortunately, two images of similar structure, but different intensity values will unlikely to have $r_{fg} = 0$ and the contrast range and stability of the NDP has not been systematically established.

Fast Fourier Transform

In the FFT method, we perform a comparison in the frequency domain. This enables translation shifts in the spatial domain to be represented as phase shifts in the frequency domain. These shifts extend the cross correlation to enable comparisons with spatial offsets between the template and experimental patterns. In physical reality, these shifts can be a combination of an incorrect pattern centre and/or a misorientation. The misorientation freedom enables a greater offset in orientation between our template and experimental pattern, which in turn enables a sparser orientation sampling of our library, i.e. fewer points in SO(3) needs to be sampled. A reduction in the library size has a significant computation advantage (a smaller library and fewer cross correlation calculations).

In preparation for the FFT-based cross correlation, both images are cropped to be square[1] and the pattern centre of the cropped image is updated accordingly[2]. We then convert the images to be compared into the Fourier domain, using the FFT (Equation 4)

$$F(u,v) = \sum_{x,y} \frac{f(x,y)}{\sqrt{MN}} \exp\left[-2i\pi\left(\frac{ux}{M} + \frac{vy}{N}\right)\right].$$

Equation 4

Where capital letters $F$ and $G$ denote the Fourier transform of an image (e.g $\text{FFT}(f) = F$), $M$ and $N$ are the image dimensions in Fourier space and the FFT coordinates are ($u$, $v$).

Comparison is performed in the frequency domain, after the application of filters, windowing and upsampling of the peak to enable subpixel shift resolution, as described previously by Wilkinson et al. [17]. This process enables background correction (through a high pass filter), removal of high frequency noise (through a low pass filter) and management of alias artefacts via windowing and leaves high contrast Kikuchi patterns remaining.

Cross correlation using a 2D-FFT approach also allows freedom to measure suitable correlations if there is a horizontal and/or vertical translation of the two images (and that

---

[1] Non-square patterns can be compared but this can be computationally difficult, as the FFT combined with the filters we use is easier to employ for a 2$^n$ x 2$^n$ field of view.

[2] The PC is defined in units of screen size, with the PC in the x direction being in units of screen width. Cropping the screen to be square, thus, changes the screen width and hence the fraction of the screen used for the PC. E.g. if the screen is cropped form 200 to 128 pixels and the PC in x was 0.4 pre crop, post crop it would be 0.34.

translation can be measured from the XCF peak position). Cross correlation of the 2D FFT is performed using:

$$\frac{\partial |r_{fg}(x_0, y_0)|^2}{\partial x_0} = 2\text{Im}\left\{r_{fg}(x_0, y_0) \sum_{u,v} \frac{2\pi u}{M} F^*(u, v) \times G(u, v) \exp\left[-2i\pi \left(\frac{ux_0}{M} + \frac{vy_0}{N}\right)\right]\right\},$$

Equation 5

where,

$$r_{fg}(x_0, y_0) = \sum_{x,y} f(x, y) g^*(x - x_0, y - y_0) = \sum_{u,v} F(u, u) G^*(u, v) \exp\left[-2i\pi \left(\frac{ux_0}{M} + \frac{vy_0}{n}\right)\right].$$

Equation 6

and $f^*, g^*, F^*$ and $G^*$ represent the complex conjugates of $f, g, F$ and $G$ respectively and $(x_0, y_0)$ are the translations between the two images in the spatial domain that give the best match. This freedom of translational correlation is important when there is a difference in pattern centre between the test and the library, and when there are subtle changes in orientations (which will be explored shortly). We note that Equation 5 is a formal formulation of Equation 3 and we include both for readability and completeness.

Normalizing the FFT correlation coefficient, by the autocorrelation functions for *F* and *G*, means the two XCFs can be compared as the coefficients for both range from 0 to 1.

The correlation is performed in the Fourier domain and these correlation functions are calculated using the 2D FFT, therefore for convenience we refer to this as the "FFT method" to distinguish this from the single value normalised inner product, which is called the "NDP" (normalised dot product) to reflect the Dictionary Indexing literature.

SO(3) Library Generation

Template matching requires a library of candidate simulated templates. Ideally these are well spaced and efficiently generated. To achieve this, we use a reprojection of an EBSD pattern (the Master pattern), which is generated using the Bruker DynamicS package. The package uses Bloch waves and the reciprocity principle to determine the reflectors and to generate a dynamical simulation [2,14,16,21]. A script was then written to generate a gnomonic projection of a crystal orientation for a given screen size and position following the conventions detailed in Britton et al. [22]. Intensities on the gnomonic projections were calculated through bicubic interpolation of the master pattern, similar to the work of Wilkinson et al [23].

SO(3) Spacing

A well-spaced SO(3) sampling improves the efficiency of the template matching method, ensuring the library has a (near) uniform distribution of orientations which are equispaced. This optimises the likelihood of finding a template and it also improves the reliability of the precision obtained. To avoid duplication and speed up computation, we use the fundamental zone within SO(3) and thus the spacing of the SO(3) sampling is dependent on the crystal symmetry being used (to avoid oversampling due to symmetrically equivalent orientations). The MTEX 5.0.1 orientation generator was used to generate equispaced crystal orientations in SO(3) space. The generator does this by splitting SO(3) into $S^2 \times S^1$. MTEX samples $S^2$ (the surface of a sphere) by splitting the polar angle equidistantly and sampling the azimuthal direction as required. Any rotation about points on the sphere are captured with $S^1$, which is linearly split.

In practice, a list of orientations was generated with a target sampling within MTEX and this generates a library of orientations within the fundamental zone of the cubic crystal system within SO(3). For every point in the library, we test the disorientation between this point and all the other points in the library (excluding itself).

To ensure that this method is sufficient for robust library matching, the uniformity of this spacing was tested for cubic symmetry. We define SF as the input sampling frequency SF for the MTEX library generator. For each SF, the disorientation (the minimum angle of misorientation between two orientations) from every orientation of sampling, to every other orientation of sampling, is calculated and compared to the expected disorientation using

$$\Delta g_{AB} = \min\{S_i g_A g_B^{-1} S_j, S_i g_b g_A^{-1} S_j\}, \forall \{S_i, S_j\} \in \boldsymbol{S} \qquad \text{Equation 6}$$

Where $\Delta g_{AB}$ is the measured disorientation between the orientation matrices, $g_A$ and $g_B$, which belong to the sampling orientations in the SO(3), and $S$ is the symmetry operator (of which there are 24 for cubic crystals). The symmetry operator is included to ensure the misorientation between the two orientations is at a minimum, thus we are measuring the disorientation.

The distribution of the measured disorientation angles for $\theta = 10°$, for cubic symmetry, is shown in Figure 2A. For all orientations within the sampling, the measured disorientation angle to its closest neighbour was at most $10°$. In fewer than 20% of samples, the measured SF was significantly less. The distribution of the measured disorientations to their closest neighbour is shown in Figure 2B. The MTEX algorithm has a linear relationship between input spacing and output mode, mean and maximum misorientations for each SO(3) data set. The sampling misorientation only exceeds the input by 0.182° at most (which compares well with the alternative sampling approach used in the study by Ram et al [24]).

The utility of the sampling for our template matching method requires us to determine the minimum misorientation between the SO(3) orientations and our final pattern, such that the peak can be found and that the refinement step works.

Our testing of the SO(3) sampling is presented to demonstrate that we can sufficiently populate SO(3) space with enough (but not too many) orientations that the library with a target SF contains a peak that corresponds to an orientation close to the actual orientation of our crystal (and that it can be refined). The mode, mean and median of these distributions are provided as indicators of the distribution shape within each target bin. Formal analysis of the sampling of orientation space can be more involved, and following the advice of one of our anonymous reviewers we suggest the reader consider evaluating the gap ratio [25], Riesz energy [26] and discrepancy [27].

In Figure 2B, we also observe that number of orientations and therefore templates increases significantly with a decrease in sampling frequency ($10° = 618$, $5° = 4958$, for cubic symmetry). This prompts us to explore what the optimal sampling frequency should be.

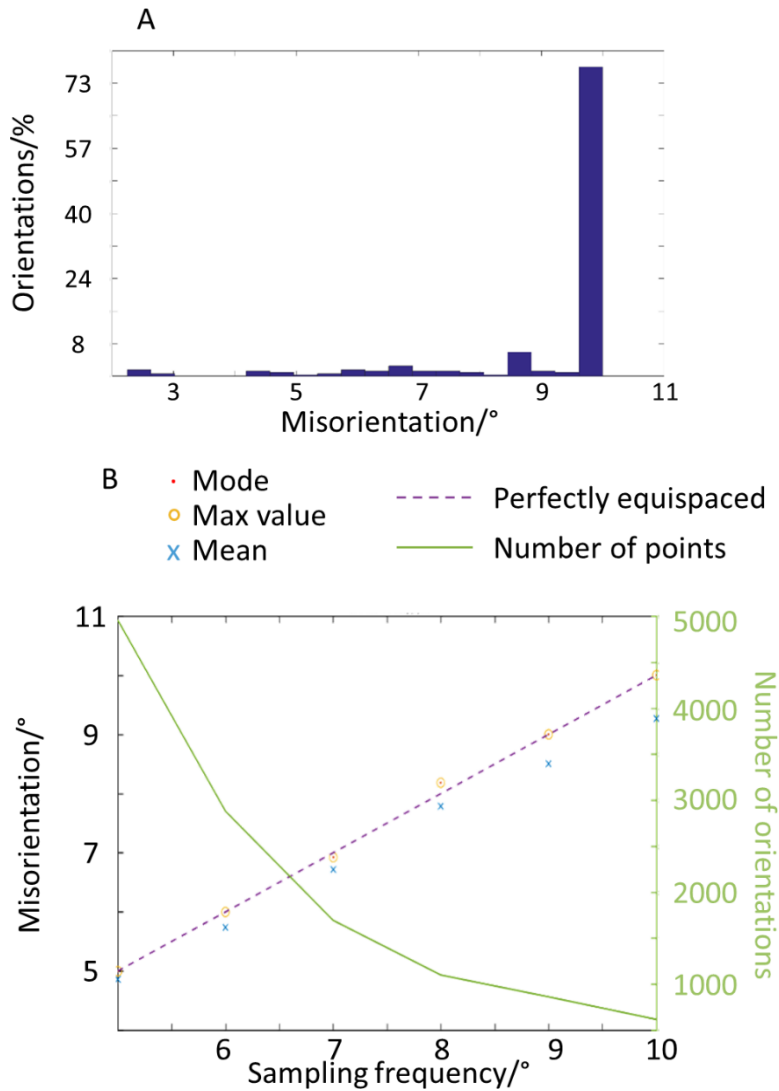

*Figure 2: A) The distribution of misorientations for a 10° sampling frequency. (B) the maximum, mean and mode distributions of similar histograms at varying sampling frequencies. Both figures are for cubic symmetry.*

SO(3) Sampling Frequency

As we have a near equispaced distribution of points in the SO(3) space, we need to work out how dense or sparse the population of orientations in the library (i.e. SF) should be spaced for suitable pattern matching. A lower SF, such as 2.5°, means that each pattern in the library is ideally misorientated by 2.5° from its nearest neighbour. As SF increases, the number of orientations in the library reduces (as per Figure 2B) increasing the size of the computations required.

In this section we want to determine the maximum SF that allows for approximate but robust determination of the pattern orientations via the template matching approach. This can be determined from a simulation-based experiment, where a reference pattern is compared to rotations of the library patterns about the three principal axes of rotation. For an example reference pattern, the pattern centre was selected to be [0.5,0.3,0.6] with a pixel resolution for each pattern of [500 x 500] pixels (descriptions of the geometry is given in Britton et al [22]). Test patterns were generated by rotating the crystal about the x, y and z axes (in the detector frame) in increments of 0.5° and sampling the same Euler angles at these rotations.

In practice, to perform the NPD, the experimental patterns are cropped, binned, and background corrected (using AstroEBSD routines, from 1600 x 1200 pixel patterns to 128 x 128 pixels) and the dynamical patterns are loaded from the master (using a cubic based reprojection in Matlab). This correction draws out the Kikuchi bands and zone axes, supressing the background signal. For simulations, it also renders the mean of the pixel intensities within each pattern to 0 and the standard deviation is fixed to 1. The templates from the master pattern have their contrast adjusted in the same manner.

For the NPD method, the correlation coefficient for the pair of patterns is calculated using Equation 1. To perform the FFT method, the experimental and library patterns are windowed, transferred to the Fourier domain using the 2D FFT, and bandpass filtered. The peak correlation coefficient is extracted from the 2D cross correlation function using Equation 5.

Template matching was performed to compare each test image with the reference image. The correlation coefficient of rotations about X, Y and Z axes are shown in Figure 3.

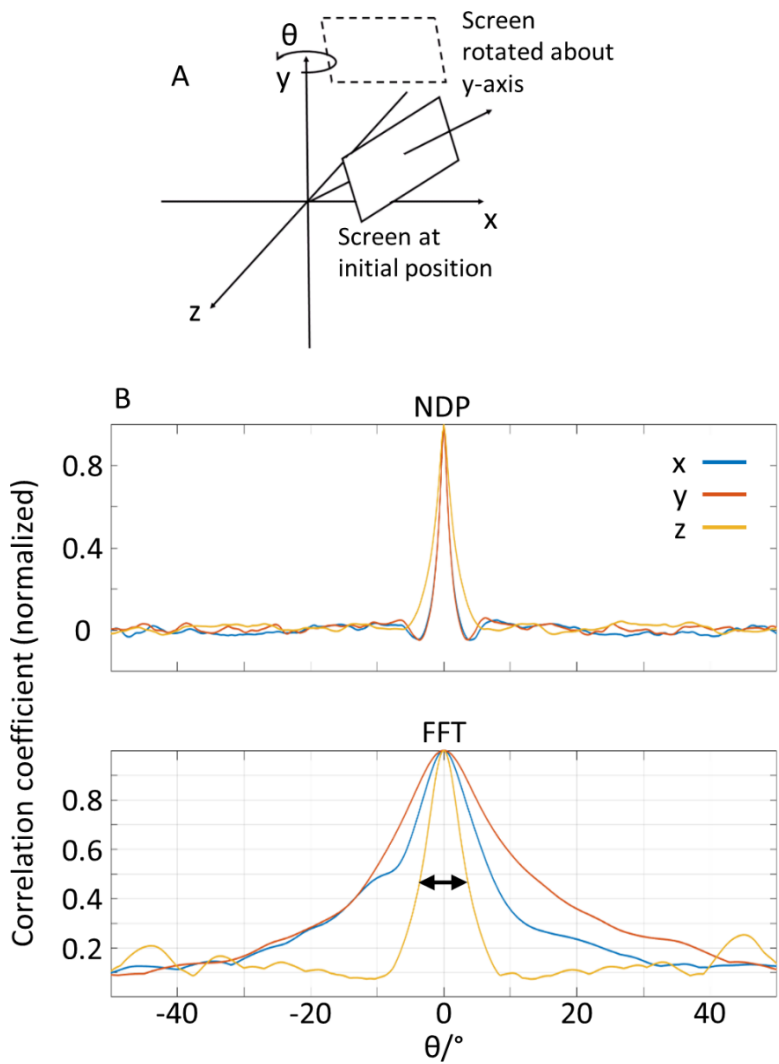

*Figure 3 A) Rotation of a generated pattern, show here for the sample y axis. B) correlation coefficients for the NDP and FFT methods from rotations about the sample axes.*

To match the experimental pattern with a template, the contrast between the correlation coefficient for the best match should be significantly higher than the baseline correlation coefficient, and therefore the best match can be identified from an incorrect match. This motivates sampling with a frequency that matches the full width half maximum (FWHM) of the narrowest correlation peak in Figure 3B.

The value of FWHM for successful matching is different for matching patterns rotated about the three axes. In practice, sampling in SO(3) space is limited by the width of the narrowest peak. In the NDP method this is a rotation about *x* or *y* and is ~2.5°. In the FFT analysis this a rotation about *z* of ~10°. The precise width of this peak will vary with screen position, crystal symmetry, and screen size.

The FFT based cross correlation includes an up-sampling of the central peak such that the peak of the cross-correlation function can be for the best translational maximum (and this is sampled with a higher orientation resolution than the SO(3) sampling). Furthermore, as we have noted, the FFT method directly enables the use of windows and filters, we can directly suppress EBSD pattern artefacts (e.g. high frequency noise and background gradients). <span style="color:red">The range of correlation values extends from 1 (near auto correlation) to 0.1 (correlation of two diffraction patterns of different orientation.</span>

For the NDP case, the maximum value will depend on the offset between the exact solution and an orientation with a misorientation of 0º (i.e. it should be close to 1, subject to small numerical fluctuations). The miscorrelation is less well defined and may not be stable.

The quality of matching can also be affected by an incorrect PC and the size of the patterns used in matching. An incorrectly defined PC can impact the correlation between two images and also result in a systematic misorientation [22]. Furthermore, if a screen has too few pixels there will be insufficient information for matching. These two factors can affect the FWHM of the peak and associated SF. We explore the sensitivity of the two methods for these problems in Figure 4 by systematically adjusting the PC position in x, y or z, shown in A, B and measuring the FWHM (from graphs similar to Figure 3).

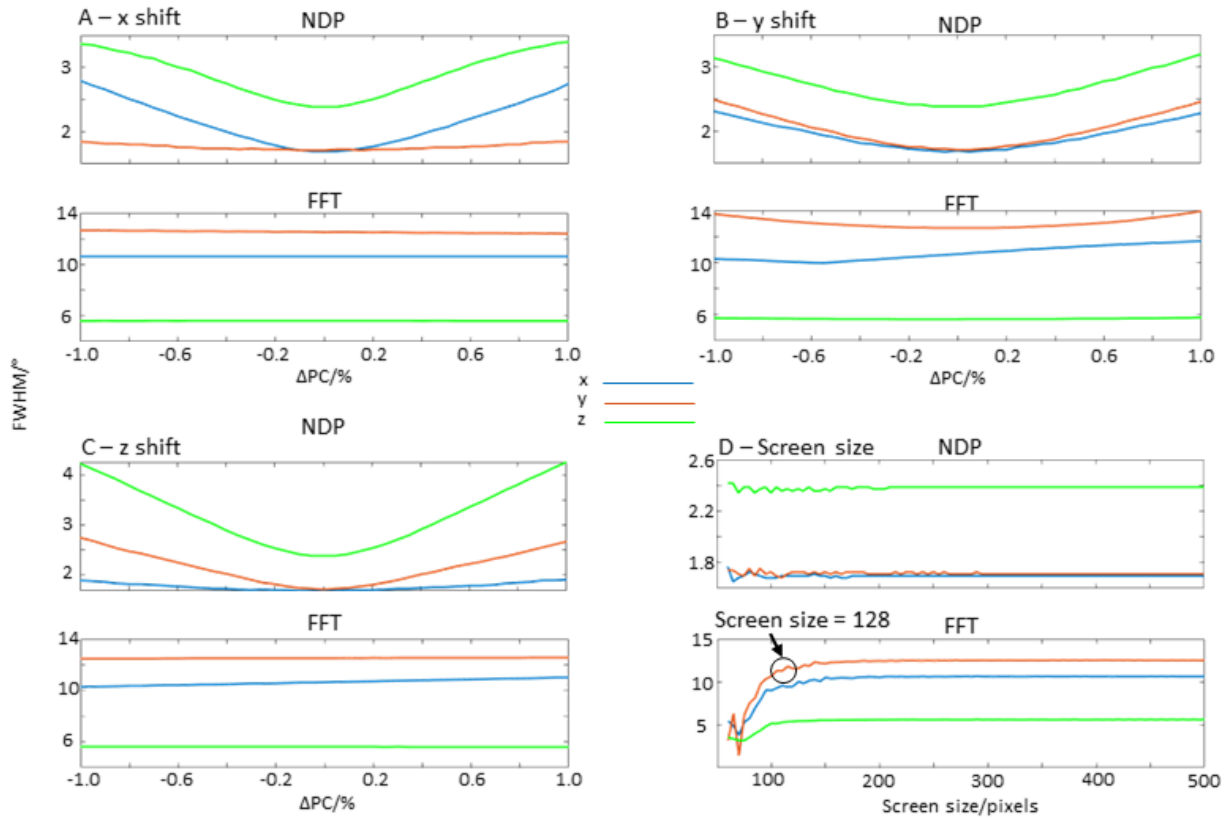

*Figure 4: A, B and C) The FWHM recorded for a shift in PC (ΔPC) in the x, y and z directions respectively. The screen is 500 x 500 pixels for each of these graphs, with the initial PC = (0.5, 0.3, 0.6). D) Varying the screen size; a screen size of 128 is labelled to identify where the FFT becomes stable for the PC set to (0.5, 0.3, 0.6).*

Figure 4 shows the FWHM for the FFT is consistently greater than that of the NDP, for both PC changes and changes in screen size, as shown in Figure 4D. For a screen size above 120 x 120 pixels, the limiting angular range of the FWHM is at a maximum of ~6°[3]. This motivates our selection of a pattern size of 128 x 128 pixels[4] (e.g. binning from 512 x 512 pixel images by a factor of 4). Based on Figure 4 A, B and C, a SF of ~2° is needed, to find a near match, for the NDP XCF and a SF of ~6° is needed for the FFT XCF. The SF stated here is slightly lower than the 10° mentioned earlier. This is based on Figure 4 showing a SF of ~6° is necessary to account for variations in PC. Note that without refinement of the orientation, a high SF would limit the ultimate angular resolution of the indexing method.

The library sampling was chosen to determine the bounds on the peak height and contrast for an allowable orientation difference between patterns in the template library and the experiment. We have performed this sampling in the detector frame of reference as this establishes a bound on the maximum allowable SF required to find a reasonable orientation from within the template library. The full width half maximum is indicative of an appropriate selection for the SF required and as the contrast of our peaks is high we can use a slightly larger SF, of 7°, in this work.

---

[3] We are using a rotation about the z axis here as Figure 4B Shows this is the smallest SF we can use to resolve the peak.
[4] This is the closest $2^n \times 2^n$ pixel screen size to 120 x 120 pixels, allowing for a more efficient FFT and more simple windowing, which is important for the refinement step.

## SO(3) Library Size and Computational Costs

In a template matching approach, searching the library for the best pattern is computationally expensive.

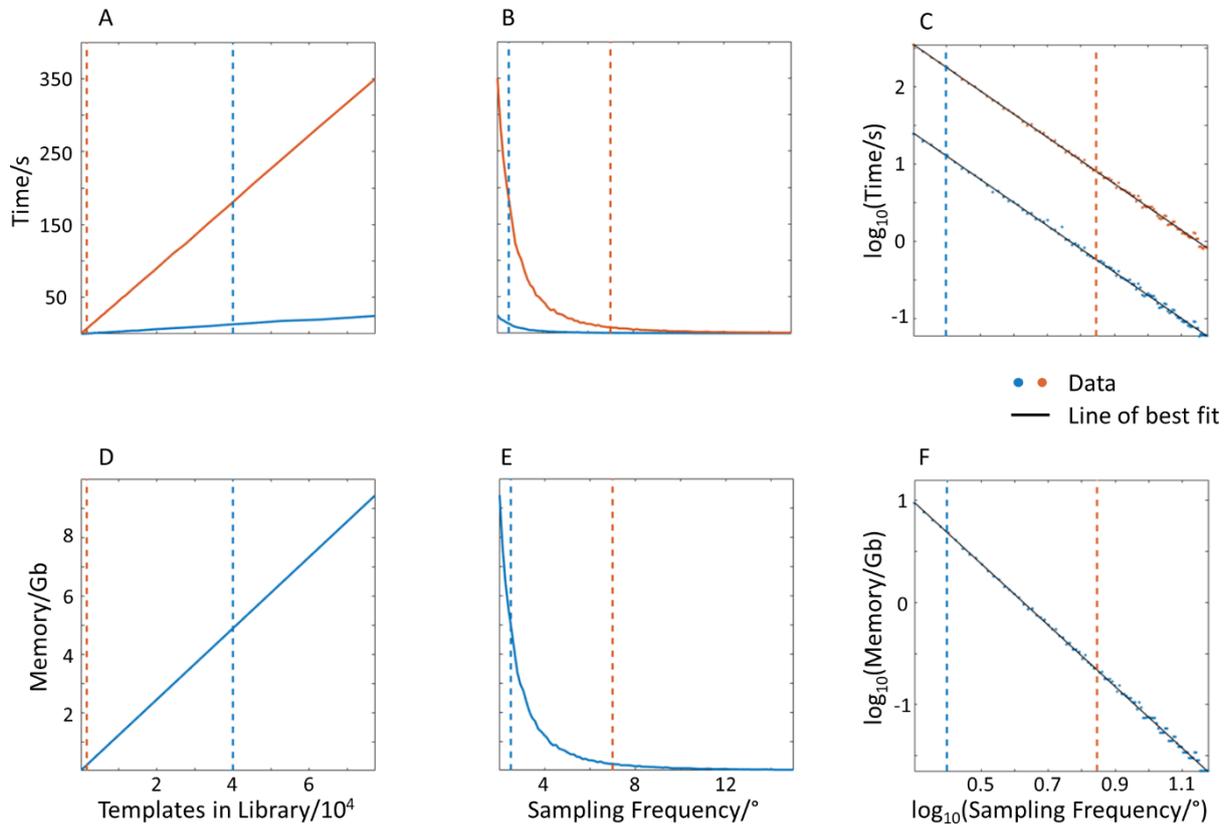

*Figure 5: The time taken to index a single pattern for the NDP (blue) and the FFT (orange) is shown in A. B shows the time taken to complete the SO(3) search. This, in turn is dependent on the number of images in the library to search through, as is shown in A, B, D and E. As the FFT uses a significantly reduced library size, it is able to index patterns significantly more quickly than the NDP. Figures C and F show the relationship between the time to index a single pattern with respect to the sampling frequency, and C the Log of the sampling frequency vs the log of the time taken. D, E and F show the same for memory usage instead of time. On all graphs, the dashed vertical lines show the respective positions the NDP and FFT libraries occupy.*

Figure 5 shows the relationship between the SF, time and memory requirements for 128 x 128 patterns and cubic symmetry. Graphs C and F all show a gradient of ~-3, which shows the time taken to index an EBSP is proportional to the SF by a factor of $10^3$. This extra cost of using the FFT for matching is therefore significantly offset due to the significant reduction in the library size, thus accounting for the speed up. It also needs less memory in the computer to store and use the library.

In practice this motivates the use of a method that can use a higher SF. It is worth noting that, although for a given library size, the time taken for the FFT to index a pattern is higher (due to costs in performing FFTs) but we note that a reduced library size (compare the intersections of the vertical lines for each method) has a significant impact on the total time required to find a match.

Orientation and Pattern Centre Refinement

We now assume that the initial library match provides us with a close approximation to the true orientation, requiring only minor refinement of the crystal orientation to obtain the true orientation. We assume the correction needed is usually within $\pm 7°$ misorientation of the crystal about the X, Y and Z axes of the EBSD detector (i.e. to within the sampling frequency of our SO(3) library).

For small rotations, we exploit the fact that there is a duality in the gnomonic projection, as a rotation about the x or y axis in the sample frame, creating a vertical or horizontal shift in the pattern (Figure 6A, provided the rotations are < ~10°). The 2D FFT cross correlation can measure this shift precisely (Equations 4 and 5). This shift can be measured using a projection of the pattern with the correct pattern centre. We can approximate this misorientation provided the PC is located within the detected diffraction pattern using basic trigonometry, see Equation 7, to correct the orientation.

$$\theta = \frac{shift}{detector\ distance}.$$  Equation 7

Refinement of the Z rotation is more difficult as it cannot be approximated by a simple translation of the pattern. It would be possible to use the High Resolution EBSD (HR-EBSD) method, described in [28] to calculate the rotation with this method but this could be computationally expensive. Instead we propose an alternative refinement where we convert the images to log polar space. The log polar transformation changes a pattern described in square [X,Y] pixel coordinates to $[\log r, \eta]$ space, where $r$ is a distance from the image centre and is the angle $\eta$ counter clockwise from the positive X axis (see Figure 6. This transformation enables rotations about the image centre to be transformed into linear shifts along the $\eta$ direction in the log-polar diagram. The log-polar transform is often used for zoom ($\log(r)$) and rotation ($\eta$) image registration. This is described in more detail elsewhere ([29,30]). We measure a crystal rotation about the Z axis between template and experimental pattern from the shift along the $\eta$ axis using FFT based cross correlation of multiple windows, extracting only the shift along $\eta$. For this transformation, we place the centre of the log-polar transform at the screen centre (placing it at the pattern centre adds significant computational complexity).

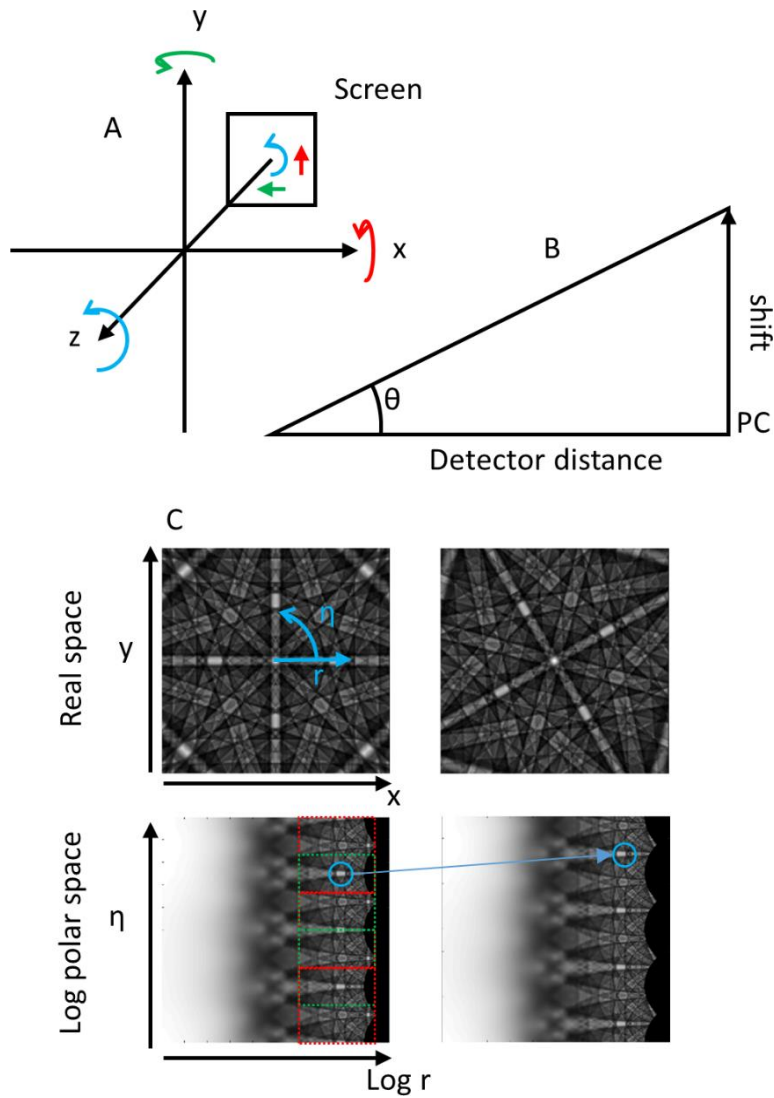

*Figure 6: A) A rotation of the crystal becomes a shift in the EBSD pattern. B) As we have the shift in x or y, or in the PC and we know the detector distance, trigonometry can be used to correct for a PC discrepancy should one occur. Finally, Z is corrected as shown in C. C) To determine the z rotation, the EBSP is transformed into log polar space, where a shift in y translates to a rotation about z in real space*

The shift used to calculate the misorientation using Equation 7 can be updated to remove the impact of the electron beam induced shift, and therefore only 1 master library needs to be generated.

The maths used in the log polar transform and conversion of the x and y shifts to rotations assumes the PC is centred on the screen, however, this is rarely the case. Fortunately, it is trivial to address this issue by iterating the correction step with the template being replaced by a pattern obtained calculated for the correct pattern centre and orientation.

The errors associated with the difference between the screen centre and the pattern centre are second order [31] and convergence to the exact solution of the measurement is demonstrated in Figure 7. The calculation is quick and computationally inexpensive (we find that ~two iterations are typically sufficient). For each refinement "step", we measure the shift in X and Y, calculate the rotation about Y and X respectively, and then perform the log-polar cross correlation for measurement of the Z rotation.

In Figure 7 we demonstrate that a first match with the closest point in our SO(3) library results in a small disorientation error between the accurate orientation and the measured orientation, as expected. A single correction of the orientation results in a closer alignment of the orientation between test and reference pattern, but there is a smaller linear error associated with the 'second order' terms (i.e. that the pattern centre is not at the image centre) and that a second correction results in an improved accuracy. This results in the ultimate disorientation between simulated pattern and measured pattern orientation which is (typically) less than 0.1°. The ultimate precision of the NDP is less than the FFT, as the refinement step benefits from the FFT based imaged shifts which are measured in the initial pattern match.

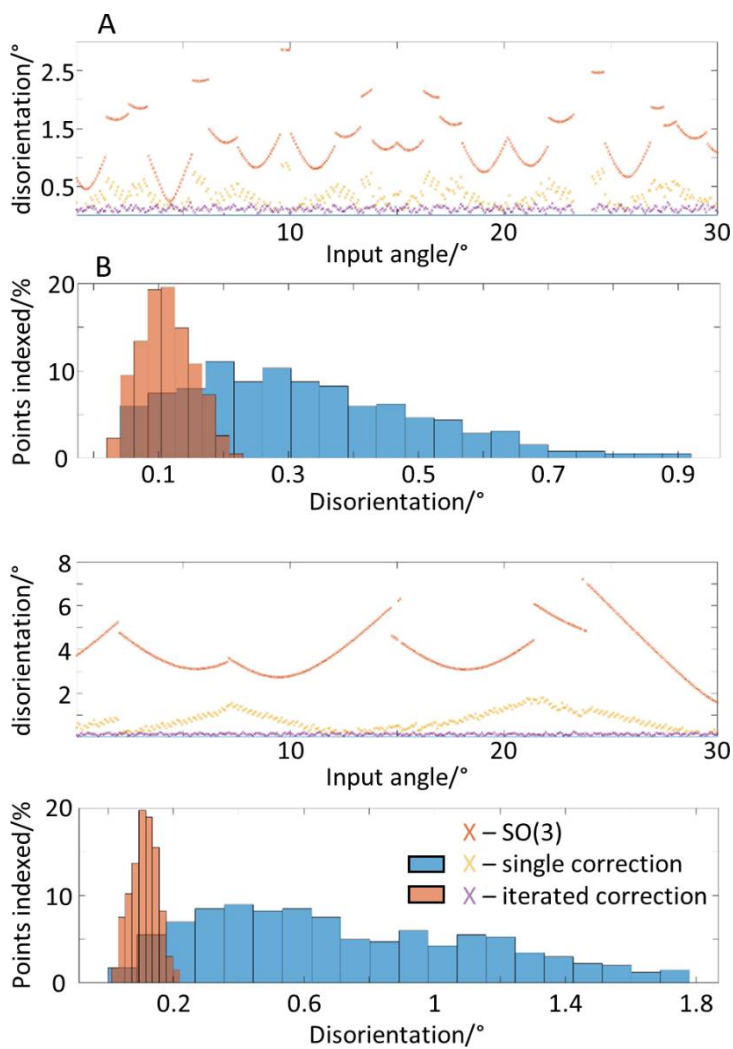

*Figure 7: A and C show the difference between the generated data and the indexed data for the NDP and the FFT respectively. They show the misorientations after the SO(3) search, a single refinement and the iterated refinement used in this method. B and D show the distribution of misorientations between the indexed and generated data. The distribution for the SO(3) search has been omitted as it makes the single and iterated distributions unreadable.*

For this test, the time taken for the NDP approach to match each pattern was significantly greater than the 2D-FFT approach as the SF is higher for the FFT (a few hours for the NDP as opposed to a few minutes for the FFT). However, this lower SF requires less correction than the FFT match, however, after two iterations of correction the results show similar accuracy between the two image comparison methods (Figure 8). It is worth noting that with each refinement, the stepping in both graphs (which is related to the SF) is reduced.

A further test of accuracy was made in which 100 randomly orientated simulated EBSPs were generated (sampled from a low SF SO(3) orientation set) and then indexed with each method using a SF of 2.5° for the NPD and 7.5° for the FFT. In both cases two refinement steps were employed to interpolate from the best fit template orientation. The misorientation between the actual and indexed orientations is shown in Figure 8 for the two methods and a similar accuracy is achieved (better than 0.1°).

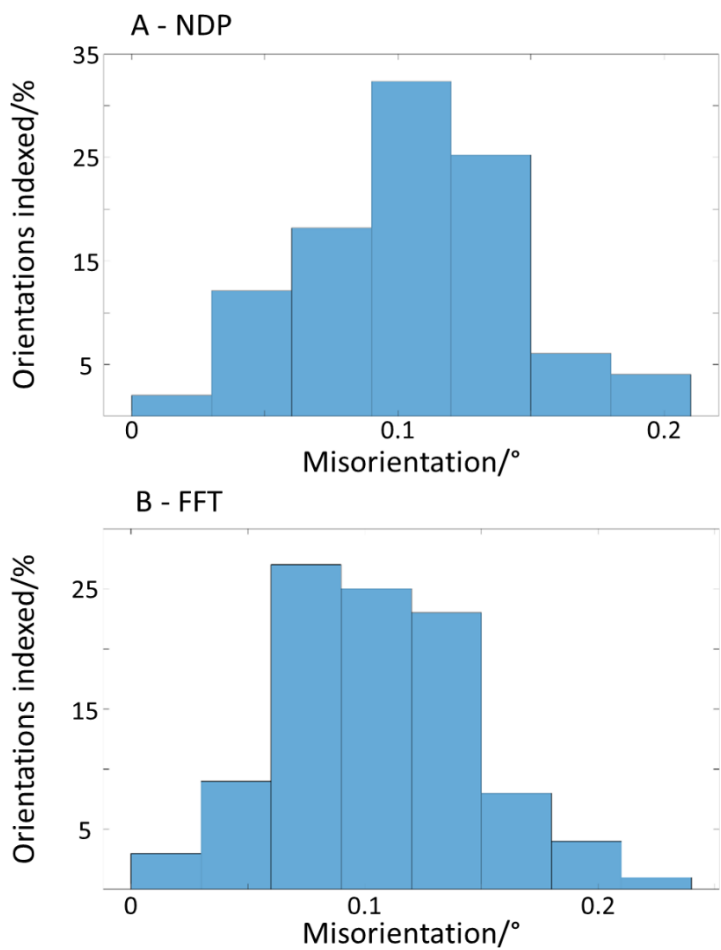

*Figure 8: The accuracy of the pattern matching approach a series of EBSPs related to a random set of orientations was generated and the misorientations of the indexed and generated patterns calculated for the NDP and FFT, A and B respectively. From B, it is seen that, using the refined FFT pattern mating approach, an accuracy or <0.2° is possible,*

Two results for the NDP indexing have had to be omitted due to a misorientation of 60° obscuring the data of a lower misorientation. Due to the misorientation it is suspected that this is a case of pseudo-symmetry (as this is cubic geometry). The FFT, in this case, seems more adept at dealing with cases of pseudo-symmetry. The methods ultimately have similar accuracy, as the refinement step is the same in both cases.

Demonstrations

Up to this point, comparisons have been made using simulated patterns. In order to demonstrate the precision of the FFT indexing method, two maps of a single crystal silicon wafer, and one map of an α-iron polycrystal deformed to 4% strain were generated. These were indexed using the refined template matching approach and compared to Hough based measurements indexed using eSprit 2.1. All maps were indexed within this software using default settings; Hough resolution of 60, and an automatic maximum of 12 bands. For these tests, the pattern centre was measured using eSprit 2.1 which employs an iterative pattern centre solver to fit the pattern centre and orientation. For these maps, the PC in the maps varies substantially across the map in both of the silicon maps.

For the template matching approach, a SF of 7° (~1600 templates), was used for the FFT, with 2 iterations of the refinement step. The NDP was not used for these examples as this paper is demonstrating the use of the new FFT method. All three demonstrations are for cubic crystals and therefore the numerical results may vary slightly if a different symmetry is present within the crystal.

Si Single Crystal:

Two maps of the Si wafer are presented, with one spanning 0.28 mm x 0.21 mm, with a step size of 10 µm and the second 3 mm x 2 mm, with a step size of 50.76 µm. These maps have a large variance in the PC due to electron beam shifts. We note that this method of precision assessment is similar to the one employed by Humphreys et al. [32].

Orientation maps in Figure 9 and 10 (parts A, B, C, & D) indicate that the sample is a single crystal and is oriented as expected (orientations are presented in the sample frame, utilising knowledge of the camera and sample tilt). Consistency of orientation assessment was performed through comparison of the Hough based indexing method and the pattern matching approach (Figure 9 & 10 part F).

Precision for each method was estimated through analysis of the misorientation with respect to a central pattern within each map and results are also shown in Figure 9 and 10 parts E & G. Histograms of the misorientation angles with respect to the central point for both Hough and XCF were also generated (Figure 9 & 10 part I).

Finally, for quality assurance the XCF peak height map for matching is also presented (Figure 9 & 10 part H) and this can be used as a filtering metric for removing suspect points, similar to the image quality in the Hough based method [33].

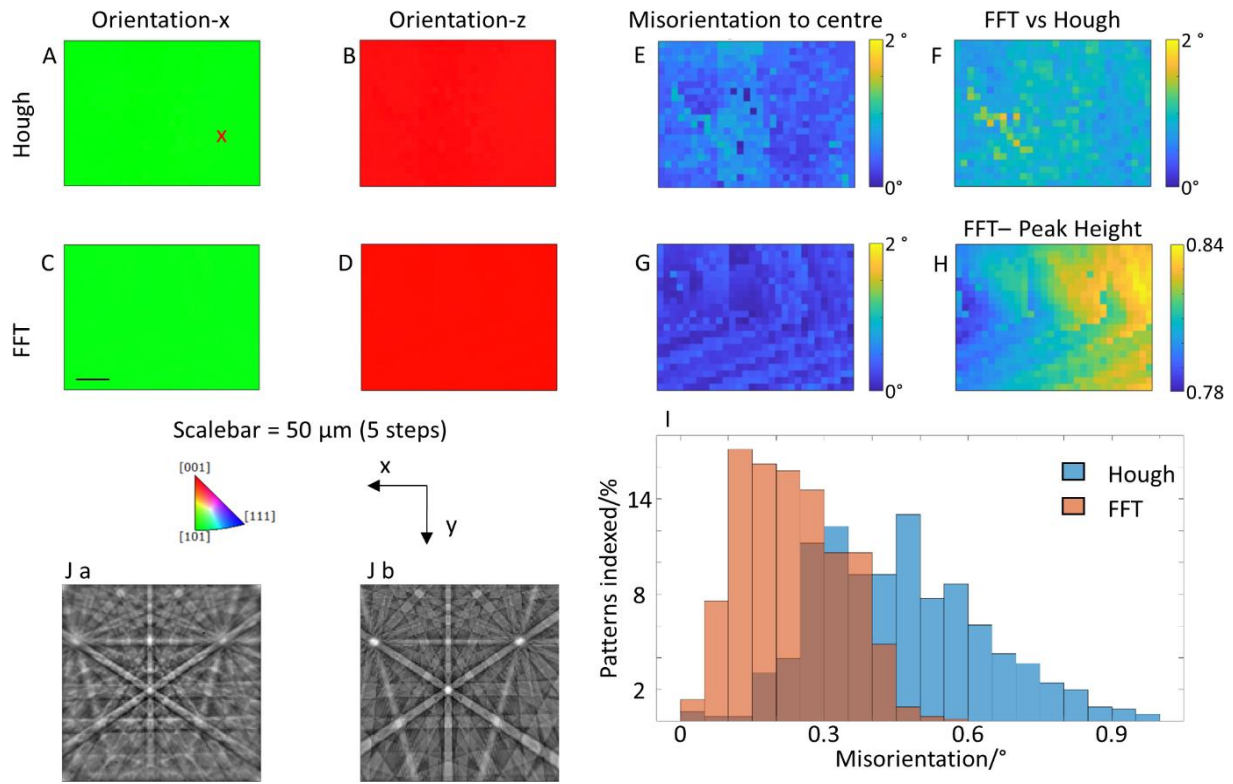

*Figure 9: Surface maps showing the orientation of small single crystal silicon for Hough (A and B) and FFT pattern matching (C and D) to demonstrate the precision of the pattern matching approach. The coloured orientation maps are shown with respect to the macroscopic X and Z axes. E and G show the misorientation with respect to the central pattern in the map. I shows the distribution of the misorientations shown in E and G. H shows the peak height from the FFT for each point in the map, this can act as an error metric similar to the image quality in the Hough. J shows the captured image, cropped and background corrected (a) and the pattern indexed with the refined method (b). The red x on the map in A shows the location from which the pattern was extracted.*

Figure 9 A, B, C and D show results from the refined template matching approach. It is shown to give similar indexing to the Hough based approach ([110] points along X and [001] points along Z). Figure 9 E and G show that the FFT method has a flatter spatial distribution of misorientations across the map, which is represented also the histogram analysis in I. Figure 9 I shows the FFT indexing has a lower average misorientation, from each point to the centre point, compared to the Hough indexing (~0.1° as opposed to ~0.4°).

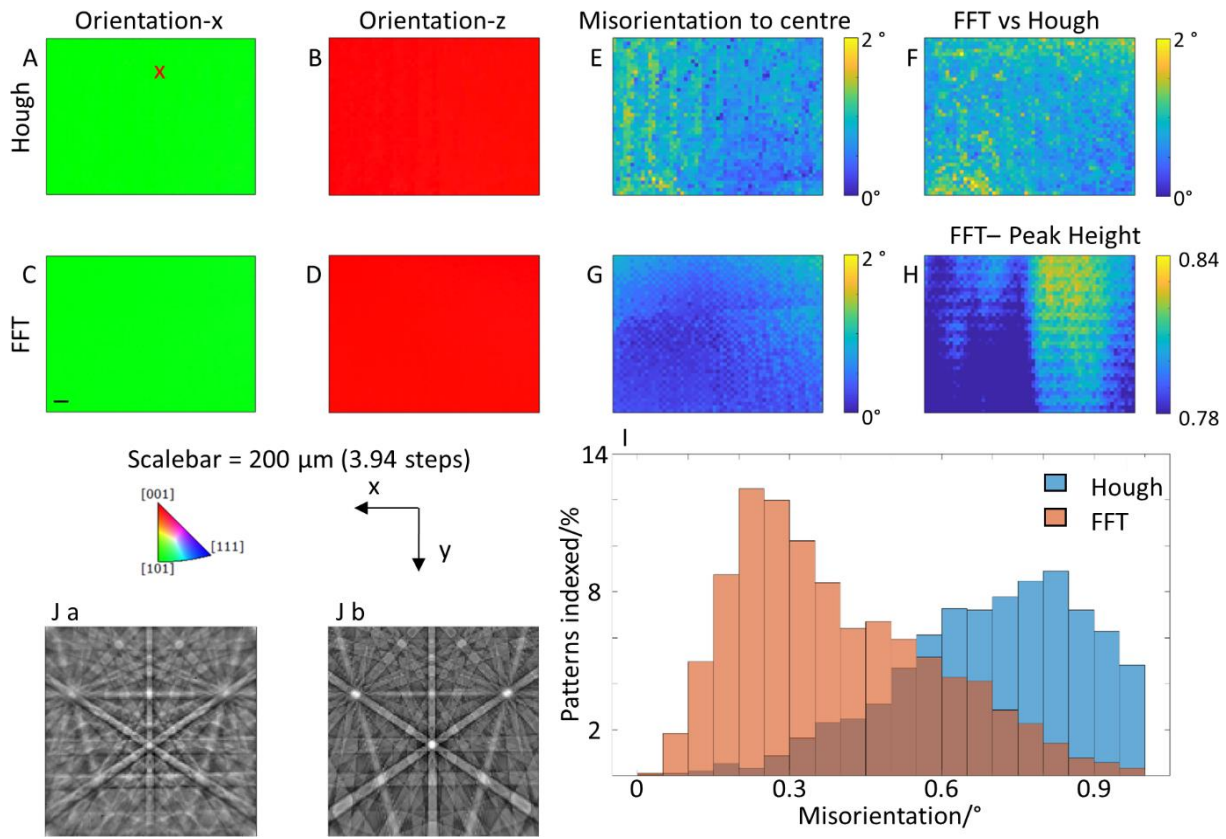

*Figure 10: Surface maps showing the orientation of small single crystal silicon for Hough (A and B) and FFT pattern matching (C and D) to demonstrate the precision of the pattern matching approach. E and G show the misorientation with respect to the central pattern in the map. I shows the distribution of the misorientations shown in E and G. H shows the peak height from the FFT for each point in the map, this can act as an error metric similar to the image quality in the Hough. J shows the captured image, cropped and background corrected (a) and the pattern indexed with the refined method (b). The red x on the map in A shows the location from which the pattern was extracted.*

Figure 10 shows results for the wider scan area, and these are broadly similar to Figure 9. There are a small number of misindexed points in the FFT indexing, which are believed to be due to large PC shift corrections (the PC varies from 0.39 to 0.52 in X and from 0.11 to 0.19 in Y).

If more complete indexing is required, we have also indexed this map through using a bespoke library (results not shown) for each point with an updated PC. This approach is not advised, however, as it significantly increases the time taken to index each EBSP (as time and memory saving is achieved through the generation of only one master library).

Figure 10 E, G and I all show the refined template matching approach has an improved precision compared to the Hough approach. Figure 10 G shows the misorientation from the central point varies radially with distance, implying that the PC variation model imported from eSprit is not quite correct for this dataset and this systematic problem is only revealed in the template matching approach.

## Deformed α-Fe Dataset

The next test is to determine the sensitivity of the pattern matching approach and to select a cut off peak height for incorrect indexing. For this, EBSD maps of an α-Fe Polycrystal were obtained from a lightly deformed sample. Patterns were acquired and indexed to show the sensitivity of the method. This map covers 16.5 by 12.3 μm at a step size of 0.15 μm and has been described in more detail in [8]. This dataset contains abrupt misorientations across grain boundaries and smaller lattice curvatures within the deformed grains. It is more useful in determining the sensitivity of the pattern matching approach and selecting a cut-off peak height for misindexing.

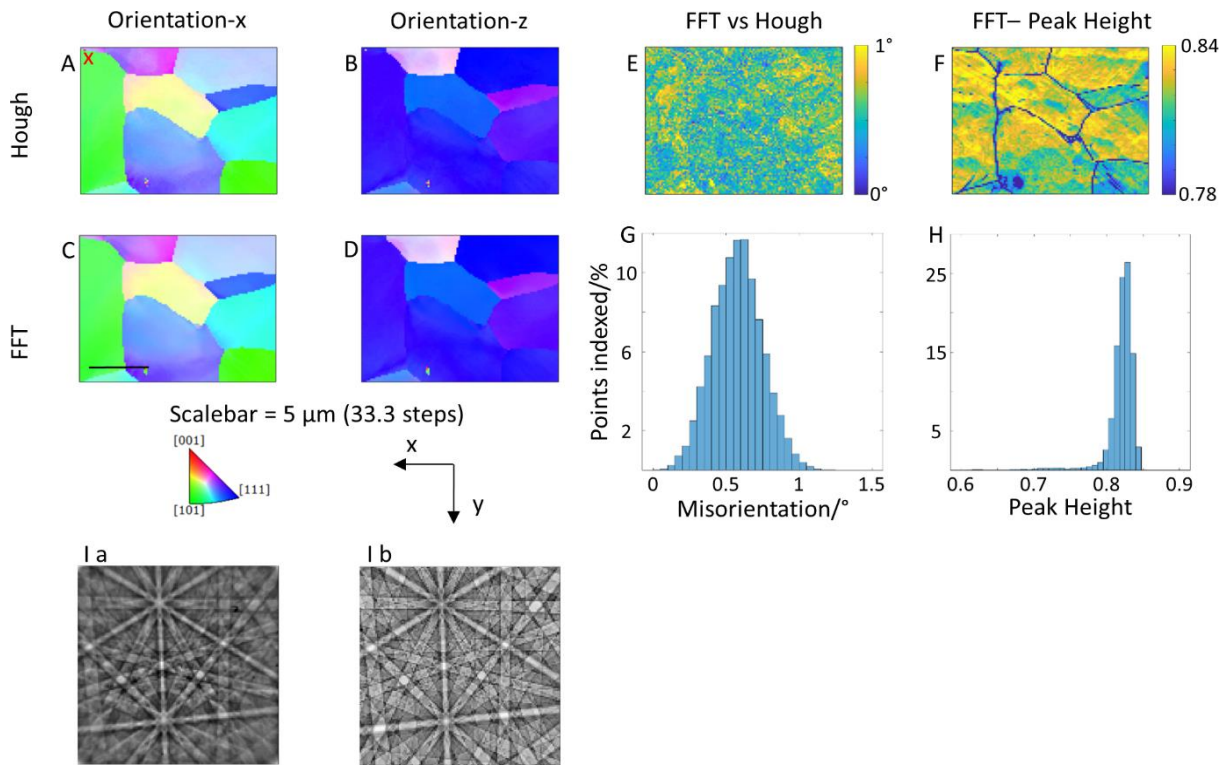

*Figure 11: Orientation maps of a deformed α-Fe sample, showing the sensitivity of the pattern matching approach (A, B, C and D). E shows the misorientation with respect to Hough as a rough guide to the accuracy of the pattern matching approach, and the distributions of the misorientations are shown in G. F shows the FFT peak heights as a map, analogous to the IQ in the Hough and H shows the distribution of the peak heights. This histogram can be used to determine where the cut off for incorrectly indexed patterns should be. I shows a) The captured image, cropped and background corrected and B the pattern indexed with the refined method. The red x on the map in A shows the location from which the pattern was extracted.*

Figure 11 A,B,C and D shows how closely the two methods index the orientations of the grains in the sample. This is shown further in E and G where the misorientation between the Hough based method and the template matching method are shown (note this illustrates precision between the methods and not accuracy in either case).

Figure F shows the peak height map for the template matching. This can work as an error metric as it reveals through low values regions where patterns may be misindexed. This map also reveals other features, such as slip bands which can be useful for deformation assessment.

The histogram in Figure 11 H, shows a high frequency of points with peak heights > 0.945 and a mode peak height of~0.955 which could be used to filter out misindexed points (such as the speck of dust near the scalebar in Figure 11F).

Discussion

We demonstrate that utilisation of the 2D-FFT for searching the template library (the master patterns) is advantageous compared to the NPD approach, as it enables a much more sparsely populated library to be used as the peak required for successful matching is wider and has higher contrast. We have not tested this method for pseudo symmetry [34] or more subtle variations (e.g. non centrosymmetric crystals [35]) which is one potential attraction of the pattern matching method in comparison to the Hough/Radon based methods. We note that in these cases, it is rare that the user does not know of the potential for pseudo symmetry problems or that the intensity variations may be more subtle, and suggest that either the refinement step could be updated (ideal, but requires new algorithm development) or a higher SF initial match could be used, following assessment of the FWHM and contrast variations associated with the likely incorrect matching.

In the FFT approach, we explicitly use image filtering to assist in drawing out the contrast associated with the orientation data. The background does contain significant information [36] but it is a complicated mix of electron channelling-in, topography, phase, and microscope conditions, and so we elect to discard this information to generate robust and high contrast FWHM and make template matching more successful. In the 2D-FFT this is also computationally in-expensive to implement, and our experience with HR-EBSD approaches [17] has proven advantageous.

For our approach to be widely attractive, we require high quality dynamical simulations and we are fortunate to have these now courtesy of the work of Winkelmann [2,14,35,37,38] and the commercial DynamicS package. Similar simulation capability has also been implemented by de Graef and co-workers [10, 11, 24] and are available within EMSoft. The master pattern approach ignores significant instrument variation (e.g. contrast associated with the take-off angle) but it is relatively robust and interpolation from the master pattern is quick and sufficient for our present needs. We have found that the stereographic projection is sufficient at present, but alternative projections [39]) could be useful particularly in optimising the accuracy of the refinement step.

The SO(3) library generation, using sampling obtained from MTEX [13] has proven sufficient in these cubic examples. Alternative sampling schemes are available (e.g. cubochloric [11]) but these may not be necessary, especially if refinement is used. For lower symmetry materials, the number of orientations in the template library will be necessarily larger. We note that in borrowing the SO(3) sampling from MTEX [13], we benefit from improvements in the MTEX open source packages to ensure that our sampling remains robust in the future use of the refined template matching approach.

The refinement step we utilise here is well suited towards 'conventional' EBSD geometries, as the solution converges well and the difference in screen centre and pattern centre makes the orientation updates converge well and we benefit from the duality of the gnomonic projection and crystal rotations. We have tried this refinement for an off axis TKD geometry [40] and this is no longer true, and so the refinement step will need to be adjusted. However, our preliminary assessment did find that the FWHM of the XCF method is still greater than the NDP method and so there is potential benefit even if a coarser refinement method is employed (e.g. use of a sampling grid or more sophisticated algorithm such as a genetic search).

A sparse set of templates limits the ultimate resolution of the technique, unless a refinement step is implemented. Refinement to the output orientation is performed using a fast-iterative approach that exploits the fact that rotations about the sample x and y axes can be approximated as vertical and horizontal shifts in the EBSPs, while rotations about the surface normal can be obtained from a shift along the angular axis of the log polar transform of the EBSPs. This interpolative refinement achieves an ultimate orientation accuracy that is dramatically superior to the initial discrete sampling of the SO(3) space. The proposed new approach has been compared to the conventional EBSD patterning indexing (commercially available) Hough based approach and with an existing "dictionary-indexing" method.

The reduction in library size is significant (a factor of 1/25 for cubic symmetry) and results in a speeding up of the indexing and a reduction in the memory overhead. The benefit on memory requirements is simple linear scaling, for a screen size of 128 x 128 pixels, the memory requirements for the NDP is ~5 Gb, and 0.2 Gb for the FFT. For computational times the gain of factor of 1/25 from reducing the library size is offset slightly by the factor ~3 increase in calculation times per pattern required to undertake the more demanding 2D-FFT analysis (see Figure 5). Two iterations of the refinement corrections incur an additional overhead of 0.03 Gb per iteration, the largest additional contribution to this is the calculation of the log polar transform. If the ~0.1° accuracy demonstrated here for the new 2D-FFT based analysis were to be targeted for delivery solely through pattern matching to a discrete template library then the SF of ~3.5° and would require a much larger library of 14,232 patterns (compare with 1,600 patterns for 7° and cubic symmetry, i.e. 9x faster)

Table 1 details the time and memory requirements needed to index the three maps documented in the Demonstrations section. The apparent similarity in the memory overhead comes from the fact that the library size is the main memory overhead in each indexing. As the library is generated once, the number of images does not affect the memory overhead significantly as storing orientations has relatively low memory costs. Indexing used a 2x Intel Xeon E5-2660 v3 @ 2.60 GHz (i.e. 20 cores). MatLab 2017a using Windows 10.

*Table 1: The time and memory requirements and number of patterns indexed in each of the demonstrative maps.*

| Map | Time taken | Memory requirements | Number of patterns indexed |
| --- | --- | --- | --- |
| Si – Small map – Figure 9 | ~ 10 mins | ~ 0.33 Gb | 660 |
| Si – Large map – Figure 10 | ~ 1 hour | ~ 0.33 Gb | 2596 |
| Fe – Figure 11 | ~ 3.5 hours | ~ 0.33 Gb | 9130 |

It is likely that this refinement approach can be used in conjunction with any indexing approach, e.g. Hough based analysis, and as we have noted there is further potential to include testing of different phases and pseudo symmetry using *a priori* knowledge of the phase and orientation relationships.

This method of indexing, although slower, is comparable to a Hough based method in accuracy, precision and sensitivity, in some cases exceeding it. This paper does not claim that this is the ultimate accuracy or speed possible for this method. However, it has demonstrated at least this accuracy, precision and sensitivity using a combination of simulated data, silicon single crystal data and a deformed α-iron sample.

The computational time for the generation of the Library at a SF of 7°, irrespective of the number of cores used in the computer is approximately 10 s. The time taken to generate the library increases rapidly as the SF approaches 0, however, as we use a SF of 7° in this paper we neglect the library generation time. Based on timed runs of EBSD maps, the approximate time, *t*, for a map to run, depending on the number of cores, *c*, and the number of patterns to index, *n* is:

$$t \sim \frac{32}{c} n,$$

Equation 7

For cubic symmetry.

## Conclusion

This paper has presented a pattern matching based algorithm for EBSD pattern indexing. The approach uses a fast Fourier transform-cross correlation function, which requires a significantly smaller library to be used to fully represent the SO(3) orientations. This results in a significant speed up in indexing compared to a normalised dot product approach. The use of a refinement step is attractive to speed up the method as it allows for a sparse library without loss of accuracy. The accuracy possible from this technique has been shown to be <0.2°, for 128x128 pixel EBSPs captured in a conventional geometry. This has been demonstrated on two example data sets from a Si single crystal and a deformed sample of α-Fe. The method is shown to be invariant to PC differences and sensitive enough to identify subtle changes in orientation, within a grain.

## Acknowledgements

The authors would like to acknowledge Jim Hickey for his contribution of the iron dataset used in this paper (which is available via https://zenodo.org/record/1214829). We thank Aimo Winkelmann, whose discussions led to the development of the PC correction step and, improved robustness in the pattern matching approach. Experimental EBSD data was collected in the Harvey Flower EM suite at Imperial College London, on an SEM funded by The Shell AIMS UTC. TBB acknowledges funding of his fellowship from RAEng. TBB, DC and AJW acknowledge funding from EPSRC through HexMat (EP/K034332/1). AF is funded through an EPSRC studentship. We thank Katherine Lumley for supporting preliminary work on pattern matching. We thank our peer reviewers for their significant help in improving the present work.

## Author Contributions

AF developed the main algorithm, with advice from the team, following preliminary work by DC, TBB and AJW. TBB collected the Si data. AF drafted the manuscript. All authors have contributed to the final manuscript.

Data Statement

High resolution figures are contained in a supplementary zip file (DOI to be inserted at paper acceptance). The silicon maps will be released via zenodo (DOI to be inserted at paper acceptance). The Fe data set is available at https://zenodo.org/record/1214829